\documentstyle[psfig]{article}

\def\farcs{\hbox{$.\!\!^{\prime\prime}$}}

\begin{document}

\begin{center}

Analysis and Calibration of Natural Guide Star Adaptive Optics Data

\vspace{0.5cm}

Eric Tessier

\vspace{0.5cm}

Royal Greenwich Observatory, Madingley Road, 

Cambridge, U.K. \ CB3 0EZ

\vspace{1.0cm}
{\large\bf ABSTRACT.}
\end{center}

From natural guide star adaptive optics data taken 
with the Come-On Plus  and with the Starfire Optical Range 
Generation II instruments in the JHK bands and in the I band respectively, 
we describe  and analyse the point spread function.  
The ultimate exploitation of  adaptive optics images requires the 
deconvolution and therefore the  calibration of the point spread function 
which is commonly made by observing a point source close to the astronomical 
target. In the partial  correction  regime, the calibration mismatch which is 
the main  source of noise or bias in the deconvolution process is induced 
by the varying seeing conditions. 
We therefore stress the procedure  enable to increase the quality of the 
calibration and discuss  the   typical performances in term  of  astrometry, 
photometry and  dynamic range   to be possibly extracted from current 
adaptive optics images as a   function of the Strehl ratio   achieved and  
the stability of  the point  spread function. Alternative techniques
to the point spread function  calibration and other problems like 
anisoplanaticism are briefly reviewed.


\begin{center}
{\large\bf 1. NATURAL GUIDE STAR ADAPTIVE OPTICS CONCEPT}
\end{center} 

{\bf 1.1 The guide source}

We call the guide source the source used by the wavefront sensor (WFS) 
to analyse the perturbed wavefront and to 
apply  in real time corresponding commands to the deformable mirror. 
The guide source could be natural or artificial. 
Most of Adaptive Optics (AO) experiments or projects 
provide only the Natural Guide Star (NGS) mode which allows to observe
many sources in the infrared.  On the contrary, the sky coverage of the 
NGS mode is poor in the optical and the laser beacons are the  current 
technique under  experiment to provide artificial guide source at the visible
wavelengths.

Depending of the astronomical targets, the natural guide source will be:

\noindent 
i) the source itself for compact sources like close binaries or stars,

ii) a component of the source for extended sources (e.g. the nucleus 
for a galaxy or a star for a cluster of stars), 

iii) the nearest possibly available projected over the sky source 
(e.g. a star) if the source is not bright enough for the WFS.

{\bf 1.2 The point spread function} 

The image obtained in AO is that of the source convolved with the 
instrumental Point Spread Function (PSF).  The shape of the PSF could be 
described as a diffraction-limited core superimposed
on a residual halo with a size corresponding approximatively 
to the seeing disk. This halo comes from the high-order Zernike modes 
not corrected  by the Adaptive Optics system. The diffraction-limited core 
gets wider (we can represent this by a Gaussian convolution) 
when errors of correction for the low-order Zernike modes are important. 
As turbulence conditions can vary on short and long  timescales, changes in 
correction are observed.  For instance, the AO correction will be poorer 
(a wider PSF and/or a stronger halo) for low $r_0$ or fast $t_0$.  
Because of angular anisoplanaticism, the PSF is  not space-invariant.
The off-axis performance of the AO system degrades with the angular 
distance from the guide source, until it is negligible for a 
distance of the order of the isoplanatic angle $\theta_0$.
To summarize, for a given guide source magnitude and AO system, 
the on-axis PSF mainly depends on seeing parameters $t_0$ and $r_0$. 
The off-axis PSF  depends
in addition on the vertical  distribution of the turbulence via the
isoplanatic patch size.  
Finally, we get a variable PSF which is a function of time and the angular
distance to the guide source. Here is the formation of images in AO:

\begin{equation}
\label{eq:1}
I = O \ast PSF (t, \theta) 
\end{equation}

A current access to the observational parameters $r_0,t_0,\theta_0$ 
would be very useful to the study of image quality and to predict 
the level of variability of the PSF  in  equation \ref{eq:1}.

{\bf 1.3 Blind and classic deconvolution} 

Raw AO images are already informative and the information 
quality increases with the degree of correction achieved 
 usually measured in term of Strehl ratio (SR) (see Annexe 1). 
The PSF is usually  identified to the  pure diffraction-limited 
function  for a Strehl ratio $\geq 0.8$. This  total correction 
occurs under seldom circumstances and  most of the times, 
AO systems do run in the partial correction regime with the
presence of a residual seeing halo and/or a widened core. 
Consequently, the deconvolution  is necessary to determine the reality of 
some structures in the AO  image: do they come from the source or from the 
PSF?  We review few
techniques to deconvolve AO images. Classic deconvolution
algorithm requires the exact knowledge of the PSF. The next section
study the calibration of the PSF.  In anycase, the PSF calibration allows to 
estimate the static aberration unseen by the WFS. Blind deconvolution 
algorithms have been proposed to get around the calibration of the
PSF.  Blind deconvolution can achieve good results provide that the 
signal to noise ratio is high enough and the structures of the source
clearly differ from the structures of the PSF (see Thiebaut \&  Conan 1994
e.g.).  Multi-framing blind  deconvolution is a promising technique to 
improve these two weak points (see Christou et al. 1994 e.g.).

{\bf 1.4 Calibration of the point spread function}

The PSF knowledge is required to apply classic deconvolution. 
Finally, the PSF in Eq. 1 depends on too many parameters so that it is 
clearly impossible to derive it analytically. An non  exhaustive list of these
parameters is the $r_0, t_0, \theta_0$, the magnitude of the guide source, 
the geometry and the type of the WFS.  
If an  isolated point source is available
in the AO image,  it could be used as a PSF estimator provide that
the angular distance to the source to be deconvolved is small enough
in front of the isoplanatic angle. Apart this seldom case, 
the strategy to face the PSF calibration problem is not
clearly defined but here is a review of two techniques to calibrate
the PSF in AO. 

{\bf 1.4.1 A point source as a posteriori PSF calibrator}

The on-axis PSF may be calibrated by observing a close
point source referred as the  PSF calibrator 
which will be its own guide source, a short 
time later or before the observation of the astronomical target.  
This ``a-posteriori'' calibration of the PSF 
on a  point source is   the standard technique  used by the observers of 
the Come-On Plus intrument at the European Southern Observatory 
(see section 2).

The drawbacks of this technique are: 

\noindent i) the PSF is  sensitive to the  seeing variations (see section 4
and 6)
and consequently the PSF may be different for the astronomical target and 
the PSF calibrator. As for speckle observations, the observer 
can switch every few minutes  between  the source and   its calibrator in  
order to sample the  temporal seeing  variations  so as to   
minimize seeing effects.  However,  many observers usually observe the 
calibrator only once  right  before or right  after  the observation of 
the source because they have the feeling to waste telescope time.  

ii) the wavefront sensor noise which affects the correction achieved and so the
PSF is sensitive to the  source shape and magnitude (see section 5). However, 
it is possible to adjust the flux coming through by using some density 
filters. 

iii)  This technique estimates only the on-axis PSF. 

{\bf 1.4.2 PSF reconstruction from the wavefront sensor data}

PSF reconstruction from the wavefront sensor residuals data recorded at 
the same time  as the AO image  with the Come-On Plus instrument 
have been investigated by Conan (1995) as a  seeing-independent calibration 
but it  is not yet available  for Come-On Plus observers. However, 
this technique has the following disadvantages:

\noindent i) it is unable  to reconstruct the residual halo 
of seeing since WFS data cut off high-Zernike orders

ii) it does not take in account the camera optics which are after the
sensing channel.

iii) the residuals are by definition  noisy.

iv) WFS data of the guide source allows to reconstruct only the on-axis PSF.

\begin{center}
{\large\bf 2. DATA}
\end{center}
 
We present here data taken with the adaptive optics instrument of the European 
Southern Observatory (ESO) called Come-On Plus (COP) and 
with the natural guide star  adaptive optics system  of  the Starfire 
Optical Range (SOR) called Generation II (Gen~II). 
Here after, data in  the JHK bands come from COP, data
in the I band come from Gen~II. Table 1 presents the diffraction-limited 
size associated and the Nyquist over image scale sampling (a value
above $\geq 1$  means over-sampling) for each band. 

The guide source is always the astronomical target and the PSF was calibrated 
with the method described in 1.4.1. Sources are bright so that continuous 
short exposure time were often used. This has the advantage to monitor the AO 
correction. Moroever, there is no photon starvation in the WFS and in the 
AO images, photon, detector and background noises are negligible in front of 
the level of variability detected in the PSF later on (sections 4 and 6).

{\bf 2.1 Starfire Optical Range Gen~II}

The Gen~II instrument (Fugate et al. 1994) 
operates at the SOR 1.5~meter telescope 
facility located near Abulquerque, New-Mexico,  USA and delivers AO 
corrected images in the I-band ($0.88\mu m$). Gen~II provides a laser 
guide star and a natural guide modes.  It   uses a Shack-Hartmann 
WFS  which analyses the light in the R band. Number of actuators are 
241, and the close-loop control band is 143 Hz. 

B. Ellerbroek and J. Christou provided a set of NGS Gen~II observations 
carried out on December 15 1994 between 5:35 UT and 6:35 UT 
under a poor seeing of  $1\farcs54$ and a $t_0$ of 2\ milli-second 
at $0.88\mu m$ (it was winter-time at Alburquerque).  Four binary bright stars 
with separations of $0\farcs13,\ 0\farcs67,\ 1\farcs75,\ 2\farcs 42$ were 
selected to  show anisoplanaticism effects.  For each binary, we have 
five frames of 4\ seconds exposure time. Five  500\ milli-seconds frames of 
a  bright star as  a PSF calibrator are  also available but at a 
different elevation. For each star, the observation were carried out within 
two minutes. The binary  whose separation is  expected to be $0\farcs 13$ 
was not resolved.

{\bf 2.2 Come-On Plus}

The Come-On plus instrument recently renamed ADONIS (Beuzit et al. 1994) 
is currently used at   the focus of the 3.6~meter ESO telescope in 
Chile for NGS AO imaging in the  near-infrared ($1-5\mu m$). 
The Shack-Hartmann wavefront  sensor analyses the light in the optical. 
Number of actuators are  52, and the close-loop control band is $55~Hz$. 

For simplicity, we have selected observations of bright binary sources.
NGS AO data of sub-arcsecond  binaries  with COP have been provided by C. 
Perrier and J. Bouvier. Four binaries were observed in 
Dec. 1993 by C. Perrier  during a run where the  Strehl ratio was pretty 
poor in comparison to the  typical value usually achieved. 
The closest binary (0.13'') was observed 
by J. Bouvier in Jan. 1994 under  better conditions.  
The average total integration time is about 5 minutes per source and
 per band (JHK). Each time, a nearby point source was observed as the 
calibrator PSF. The  time gap between the calibration and the source 
observation is  typically 10 minutes but with some possible large  deviations 
(once 45 minutes!). 

We do not have any estimation of $r_0,\ t_0$ during these nights 
(incidently, on  ADONIS these parameters are now available for the observer). 
However, the seeing angle $\omega_0 = \lambda/r_0$ was sometimes estimated by 
the observer and ranges typically from $0\farcs 7$ to $2\farcs$ 
at $0.55\mu m$. 
The limiting factor for COP is more the coherence 
time  $t_0$ than the seeing  angle $\omega_0$. 
Good corrections under a seeing 
of $2\farcs-3\farcs$  could be  achieved 
if the coherence time is slow enough (C. Perrier, private 
communication). 

\begin{table*}
\caption{\label{tab:one}
Diffraction-limited size associated to each band and image sampling}
\begin{flushleft}
\begin{tabular}{lllll}
Instrument&Band  &$\lambda$($\mu m$) & $\frac{\lambda}{D}$ (mas)& Sampling\\
\hline
\multicolumn{1}{l}{Gen~II - SOR}\\
\multicolumn{1}{l}{ D=1.5m, linear obstruction=0.07, image scale = 35mas}\\
&I&$0.88$&$121.$&$1.725$\\
\hline
\multicolumn{3}{l}{Come-On Plus (ADONIS) - ESO }\\
\multicolumn{1}{l}{D=3.6m, linear obstruction=0.44, image scale = 50mas}\\
&J&$1.25$&$71.6$&$0.716$\\
&H&$1.65$&$94.5$&$0.945$\\
&K&$2.2$&$126.$&$1.26$
\end{tabular}
\end{flushleft}
\end{table*}

\begin{center}
{\large\bf 3. ON-AXIS PSF SHAPE}
\end{center}

{\bf 3.1 Full Width Half Maximum and Strehl ratio Diagram}

The PSF calibrator data from COP were obtained  on the 
1st January 1994 at 8:54, 8:59  and 9:03 UT respectively for the KHJ bands.
We obtained continous series of  one-second  exposure frames 
respectively 120 in the  K band and 60 in the H band and of  three 20-second 
exposure frames  in the J band.  Left plot of Fig. 1 shows the 
distribution of the full width  half maximum (FWHM) versus Strehl ratio 
(SR) for each individual  exposures of the calibrator in JHK.   The 
FWHM  is rescaled on the right by a factor $\lambda/D$ (which
corresponds to the diffraction limit at the wavelength $\lambda$ for a
telescope of diameter D), points clearly gather along a single curve.
This curve could be interpreted as the AO response  to differents
seeing conditions since $r_0$ and $t_0$ vary with the wavelength as 
$\lambda^{\frac{6}{5}}$ and have changed during the observations. 
Indeed, long  exposure (approximately   1 minute) from a set of data obtained 
during  different nights of  observation and so in different turbulence 
conditions have been overplotted.  
Again points fit pretty well the same curve. 
However, at the same FWHM, the SR seems to be a bit lower for shorter 
wavelength. This may be explained as follows: the decrease in the SR value 
comes from the widening of the  PSF core and the power in the residual seeing 
halo;  at shorter wavelength, the contribution  of the seeing halo  compare 
to the widening (or equivalently the FWHM) would be relatively higher 
due to a shorter  $t_{0}$ at equivalent $D/r_0$. 
Anyway, this curve  is the PSF response of the Come-On Plus experiment. 
Other AO systems may give different  responses (especially, some curvature 
sensor system may produce a sharp  PSF  even for low Strehl ratios).   
However, five 500ms CCD frames of PSF calibrator from Gen~II which 
were obtained  on  15 December 1994 from 5:34 UT to 5:36 UT in the I band 
are  overplotted on Fig. 1 and  roughly fit the curve.

\begin{figure}
\psfig{height=8cm,angle=-90,width=16cm,file=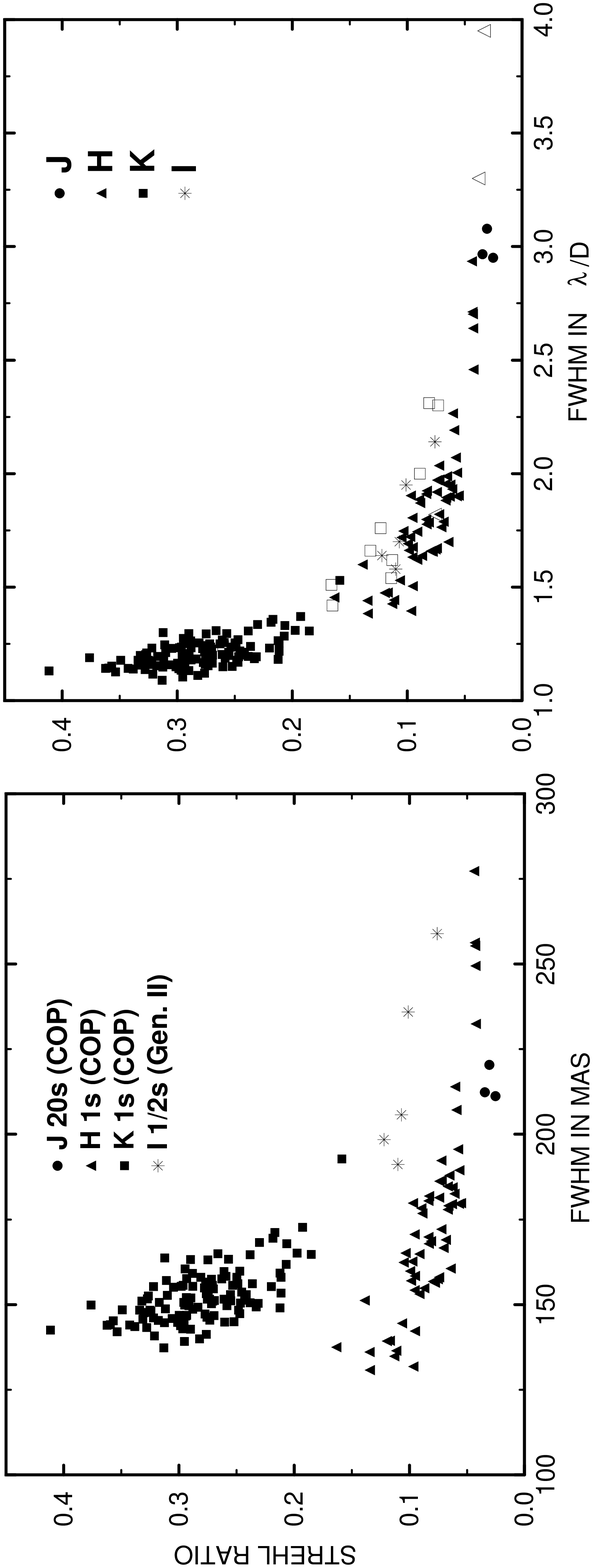}
\caption{Position of the PSF in a SR-FWHM diagram. 
Left. PSFs are one-second   and 20\ seconds  exposure images 
obtained  within 10 minutes time with  
the Adaptive Optics system  COP in HK and J respectively 
 and 500~ms exposure image taken with the  NGS AO Gen~~II in the I~band 
within 2 minutes.  Right. Same but FWHM is 
rescaled by $\lambda/D$. Long exposure 
(one minute)  PSFs   obtained with the adaptive optics system COP in HK 
during various nights are overplotted in open symbols.  See text.}
\end{figure}

As shown by this curve, the Strehl ratio describes very well the PSF with a
SR above 10\% but poorly for lower values. In the latter case, the FWHM
will provide additional information on the PSF. 
Anyway, these plots could be used to define the best strategy during AO 
observation. As a rule, the turbulence effects get worst as one goes to 
shorter wavelengths; consequently, the correction and the SR are poorer.  
However, in some cases, by going to shorter wavelengths one can get higher 
resolution in spite of a lower SR; this is because the FWHM becomes 
sharper thanks to the narrower diffraction core  (following a 
$\lambda /D$ law).  For example, in terms of FWHM, the best 
individual images are in the H-band rather than in the K-band (see Fig. 1).

\pagebreak
{\bf 50\% energy radius}

Left and right plots in Fig. 2 shows  
the 50\% energy radius R50  versus   FWHM and
R50 versus SR respectively for each PSF again.  
R50 is linearly related  to SR for SR
between 0.05 and 0.30. As the SR goes below 5\%, R50 increases steeply; for 
the visibility of the plot some extra-measurements like
(10.4,0.008), (10.7,0.007) are out the frame. Let us remind that for a pure 
Gaussian, the R50 is equal to the half of the FWHM. The R50 to FWHM ratio  
for the diffraction-limited PSF depends on the obstruction of the telescope.
The ratio increases from $\approx 0.5$ for an obstruction of 0.07 
(1.5m SOR telescope) to $\approx 0.75$ for an obstruction of 0.44 (3.6m ESO
telescope).

\begin{figure}
\psfig{height=8cm,angle=-90,width=16cm,file=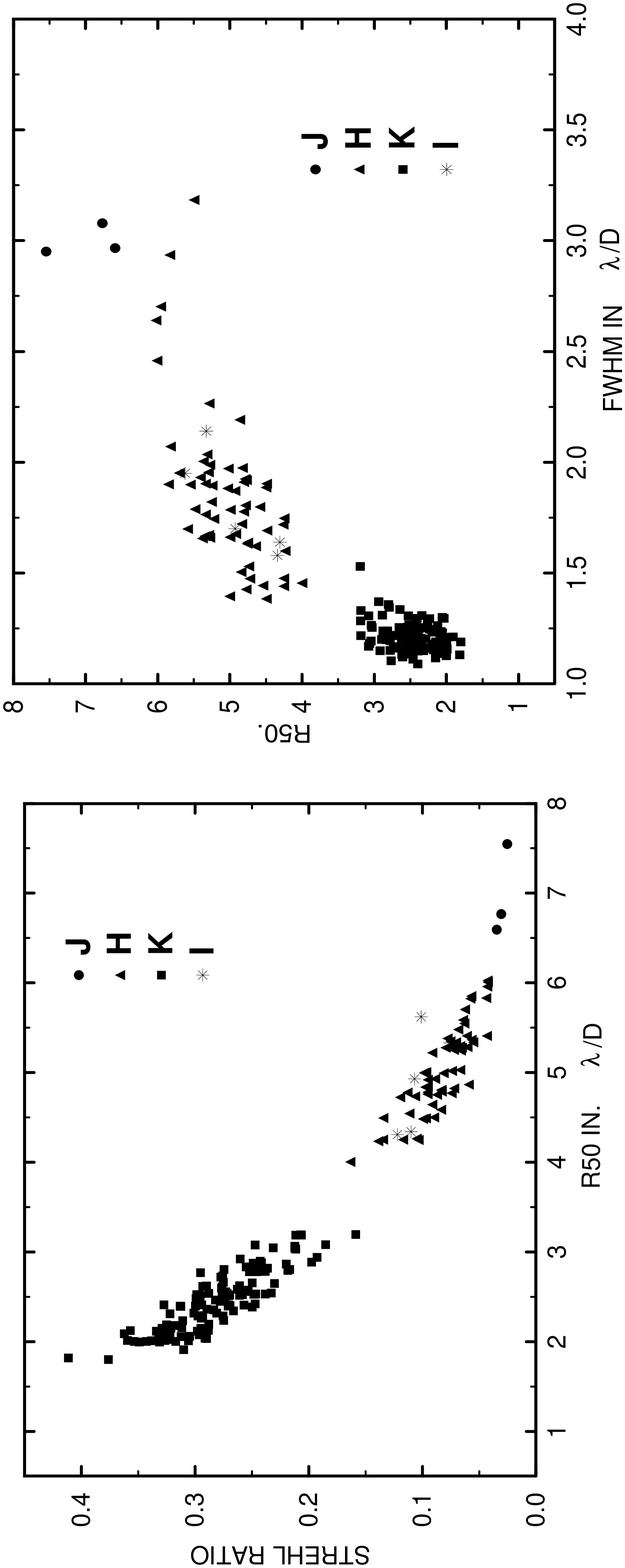}
\caption{Same as in Fig. 1 but the PSFs are located in a 
SR-R50  (left) and FWHM-R50 (right) diagrams. See text.}
\end{figure}

{\bf Power spectrum cutoff}

The spatial frequency cutoff at 40dB of the power spectrum (FCUT)
is an indication of the fall off in the power spectrum. 
That does not mean that higher spatial frequencies are unseen 
(see e.g. Fig 11). In the diffraction-limited case, FCUT is equal to 96\%. 
Left plot of Fig. 3 shows   FCUT in function of the Strehl ratio. 
The curve shape  compares well with this of  Fig. 1. In fact, FCUT  
is pretty well   
linearly related  to the  FWHM as shown  by the right plot of Fig 3.

\begin{figure}
\psfig{height=8cm,angle=-90,width=16cm,file=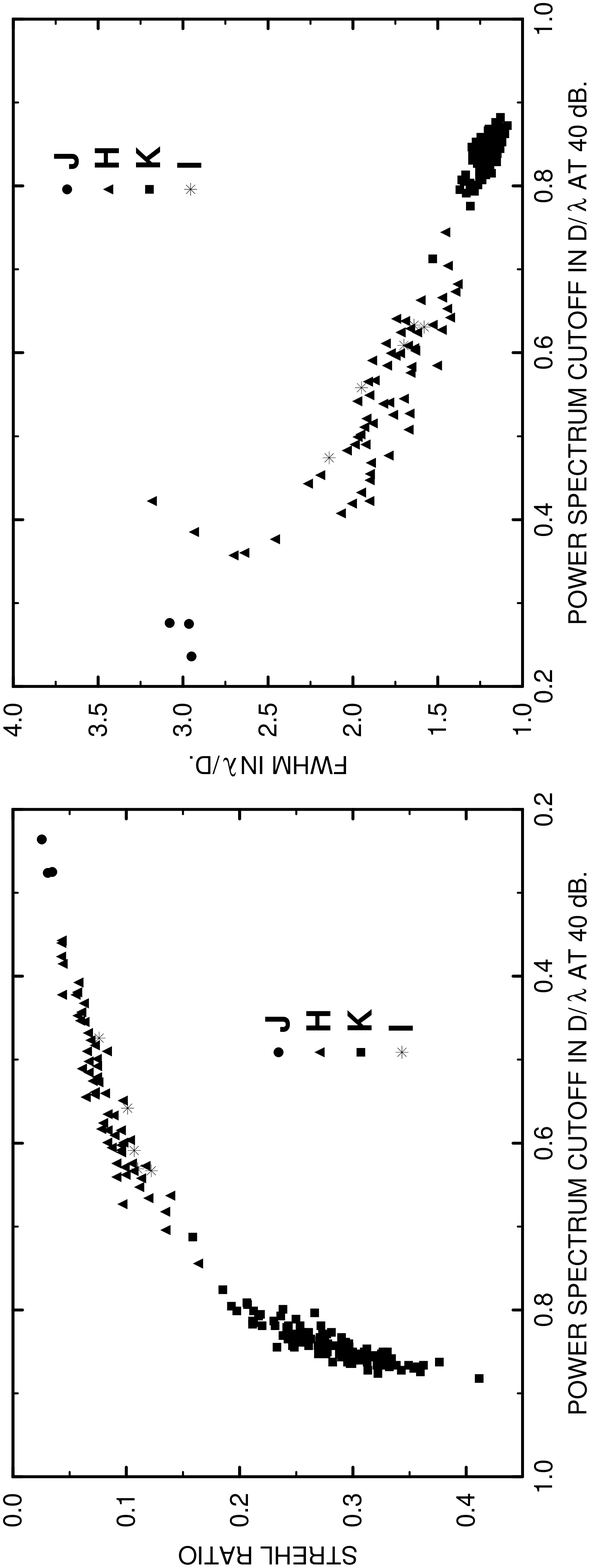}
\caption{Same as in Fig. 1 but the PSFs are  located  
in a SR-FCUT (left) and FCUT-FWHM (right) diagrams. See text.}
\end{figure}

{\bf 3.2 Profiles}

Individual exposures frames were co-addded to plot the
corresponding profiles of long-exposure PSF in Fig. 4 for different
Strehl ratios. However, the  profile of the best one\ second image in K 
with  a Strehl ratio of 0.40 compares well with the 
theoritical diffraction-limited image. The profile from 
Gen~II  comes under the comparable COP profile in the $4-12\ \lambda/D$ 
region  likely due to the higher bandwidth used. Anyway, these profiles
show than a dynamic range up to $10^4$ close to the source is possible
under good corrections.

\begin{figure}
\psfig{height=8cm,angle=-90,width=16cm,file=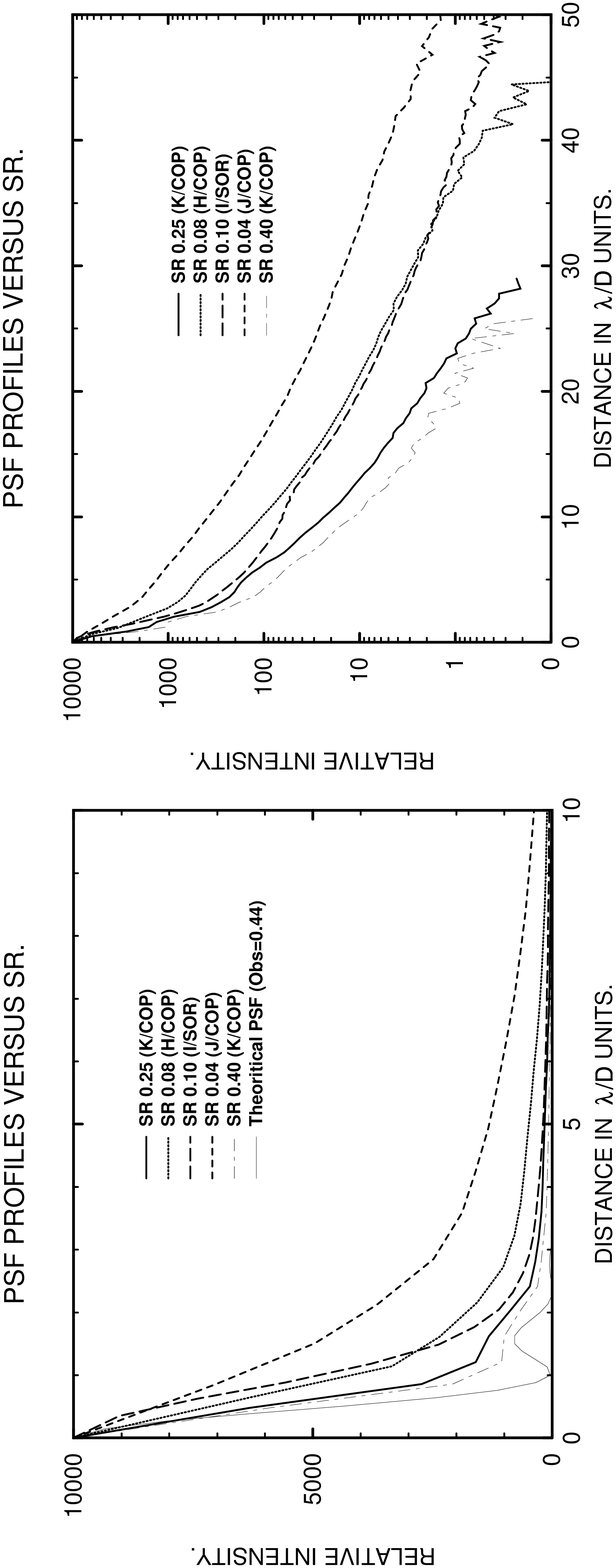}
\caption{Profiles of PSFs for different Strehl ratio from 
COP and SOR Gen~II data. 
Linear scale on the left,  logarithmic scale on the right. See text.}
\end{figure}

{\bf 3.3 Circular symmetry}

PSFs from COP  do not show in general any significant deviation from 
circular symmetry.  SOR Gen~II PSFs show a significant 
elongation  PSF (see Fig. 7),  this could come from some uncorrected 
static aberration in the imaging channel or from temporal anisoplanaticism 
 (see section 5) or  from the  jitter of the telescope.

\begin{center}
{\large\bf 4. PSF TEMPORAL STABILITY}
\end{center}

{\bf 4.1 Strehl ratio} 

SR is an excellent tracer of the PSF stability in time. As seen on Fig. 5, the
SR is highly variable in time. These variations could be induced by either 
seeing variations or  by the wavefront sensor noise  plus 
the uncorrected high Zernike terms at constant 
seeing. The answer is probably both, but seeing variations is certainly 
predominant. Some breaks or slopes superimposed to the short time scale 
variations are likely related to the seeing which is known to
vary in a comparable way. We shall see in section 6  how that affects the  
calibration and the  deconvolution process. It would be interesting to compare
the SR in time to the $r_0$ and $t_0$ measurements from a seeing monitor.

\begin{figure}
\psfig{height=16cm,angle=-90,width=16cm,file=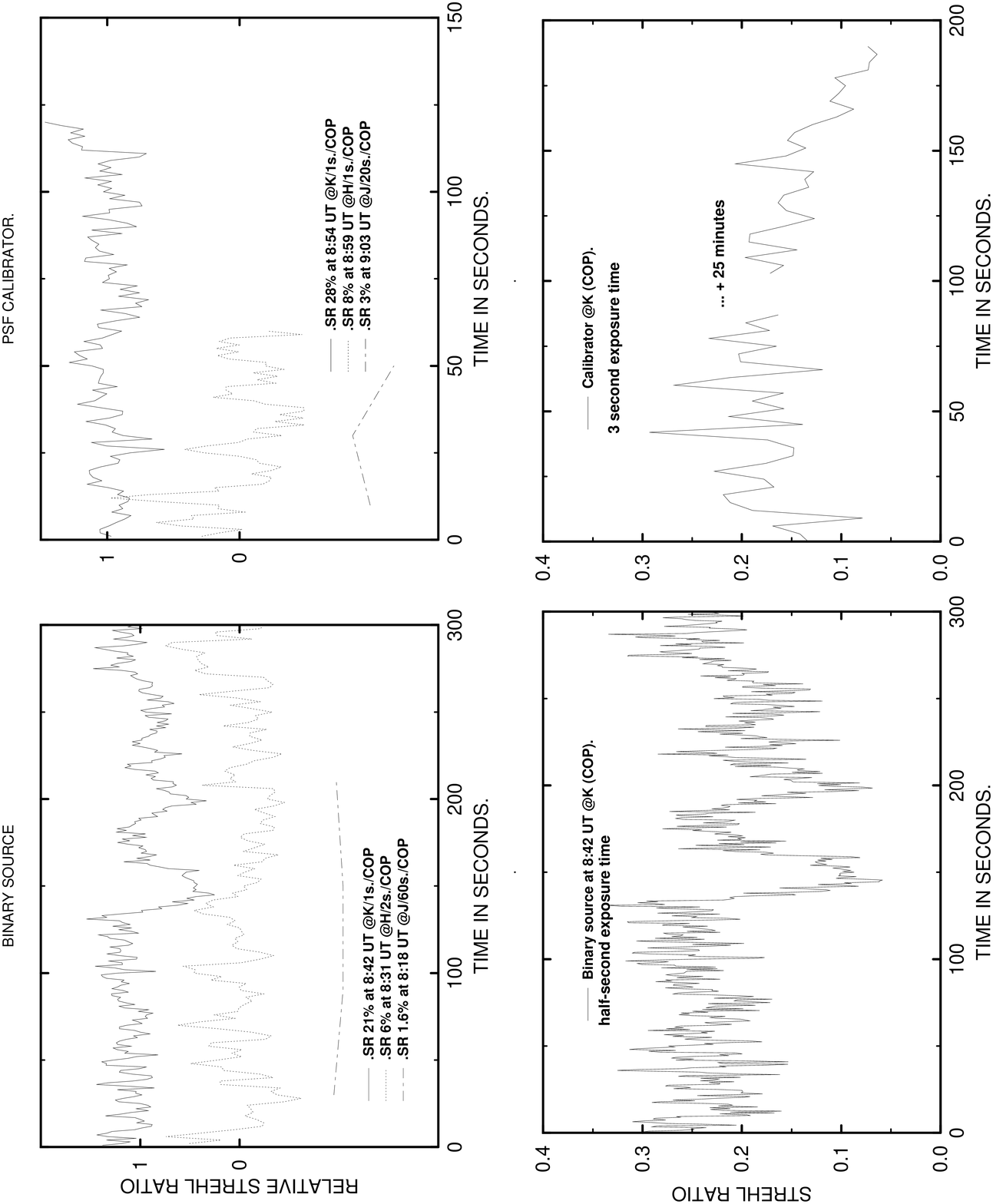}
\caption{ Strehl ratio in function of time from COP data. 
Top, for a binary source and
its calibrator at JHK. SR is normalized to its mean value.  SR is  shifted 
by -1 and -2 in HJ respectively for visibility. 
Let us remind that the Strehl ratio is a 
source-dependent value.  For a binary,  the Strehl ratio is 
divided roughly by a  factor $1+F$ where  $F$ is the flux ratio of 
the secondary to the primary. In this case, F is equal to $\approx 1/3$. 
Bottom left SR variations from  half-second exposure frames. 
Bottom right shows the evolution of the SR for a point source after a 
time gap of 25 minutes. 
See text.}
\end{figure}

{\bf 4.2 Full Width Half Maximum}

From Fig. 1, we can see that the image in the K band is very well 
stabilized, FWHM being always  less than 0\farcs 18. 
In H, the SR drops below 10\%, a key value below which the
correction is much more sensitive to turbulence effects as shown by the
large variations of the FWHM between 0\farcs 13 and 0\farcs 25.  
Again, the distribution of the points in the HK-bands illustrates how the 
PSF varies as the turbulence conditions continuously change during the 
observations. The scatter of the points depends upon the coherence time of 
the turbulence: had the coherence time been longer in that night, the PSF 
would have been much less sensitive to varying turbulence conditions.  
As a rule, the worst the turbulence, the less efficient the correction. 
The PSF variations are smoothed  with  longer exposure times  
and/or by coadding individual images.  
This is why the scatter of the 3 images in J is much reduced due to the 
exposure time of 20s compared with 1s for the HK
bands (see Fig. 1). 20\ seconds  PSF position in the  FWHM-SR and FWHM-R50 
diagrams are shown on Fig. 6.

\begin{figure}
\psfig{height=7cm,angle=-90,width=16cm,file=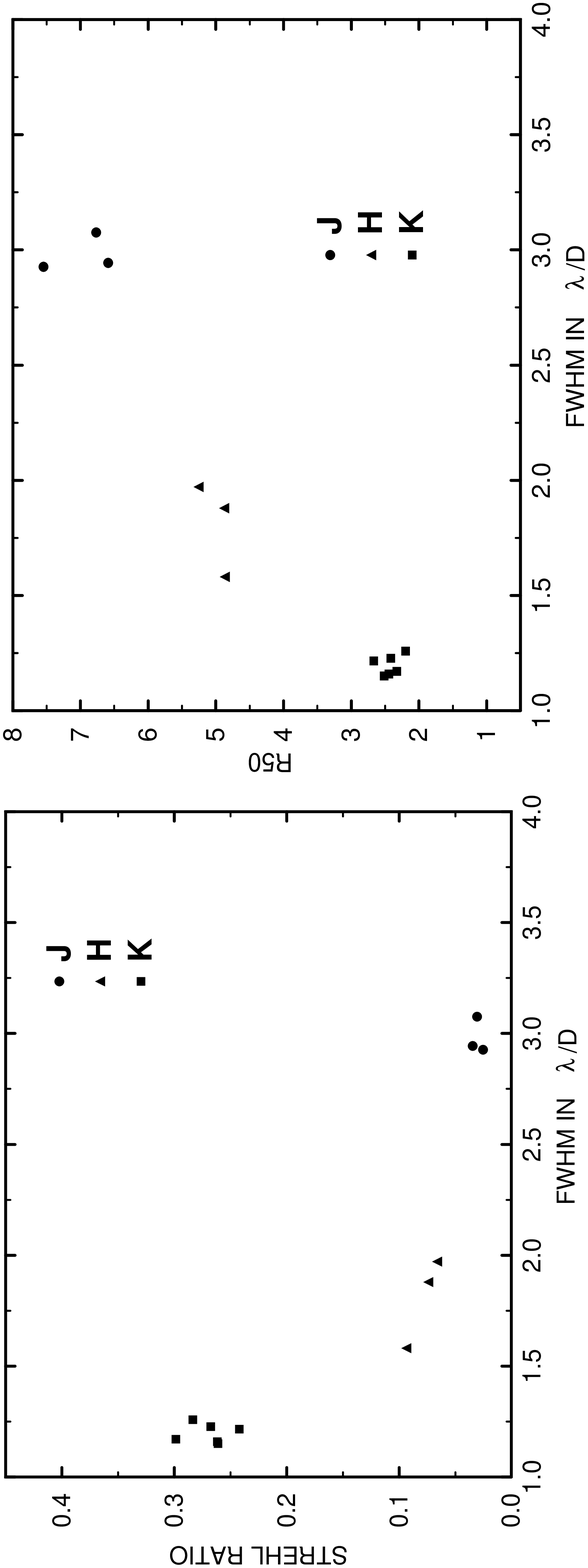}
\caption{Continous 20 seconds exposure PSF plotted in 
the R50-FWHM and SR-FWHM diagrams from COP data. See text.}
\end{figure}

Few conclusions can be derived first.

\noindent i) In the partial correction regime, the PSF is very sensitive to 
the seeing. 

ii) When calibrating the PSF, it is necessary to integrate long enough 
to sample seeing variations.

iii) As in infrared speckle interferometry,  seeing variations will likely 
limit the accuracy of the PSF calibration.

\begin{center}
{\large\bf 5. OTHER EFFECTS}
\end{center}

{\bf 5.1 Anisoplanaticism} 

Anisoplanaticism depends on the seeing conditions but let us see 
its presence on the data.

{\bf 5.1.1 Angular anisoplanaticism} 

The first effect of angular anisoplanaticism (see Wilson \& Jenkins 1994) 
is that the Strehl ratio of the PSF falls off as $\theta$ increases 
(see Eq. 1).  It is possible to use images of binaries to estimate some 
anisoplanaticism.
The on-axis PSF is obtained at the photocenter of the binary in the
wavelength used by the WFS.  For contrasted binaries, we have approximatively
the on-axis PSF for the primary component and an off-axis PSF of $\theta$ for 
the secondary component.  The second effect is that the elongation of the 
off-axis PSF in the direction  of the on-axis PSF
(see Wilson \& Jenkins 1994). 

{\bf I band} 
Anisoplanaticism effect on Gen~II data in I  was detected for the 
$2\farcs 42$ binary but  not for the $1\farcs 75$ binary. 
At $2\farcs 42$,  the Strehl ratio at about 10\%  
falls off by 25\% as shown on Fig. 7. Note that 
the PA of these two binaries are not the same.  The photocenter is at 
the zero position in each contour image. 

\begin{figure}
\psfig{angle=-90,width=18cm,file=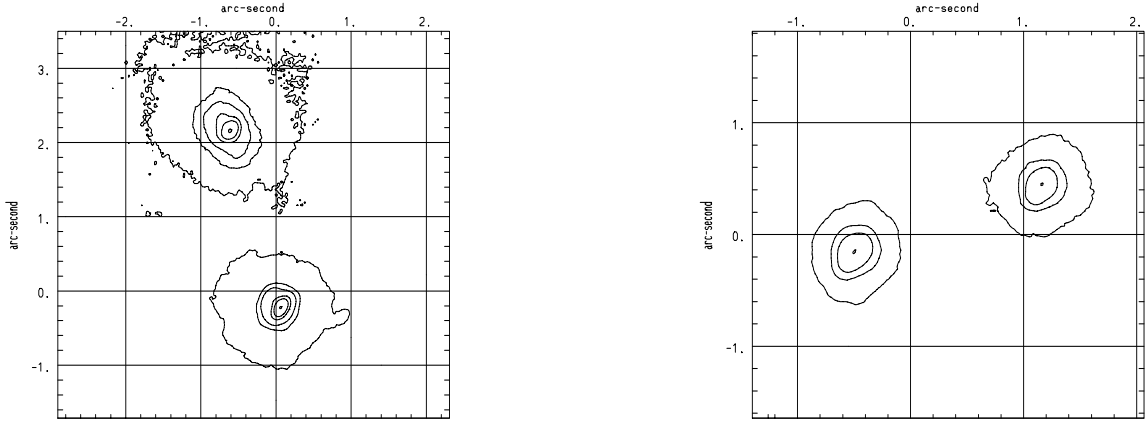}
\caption{Contour of binaries observed with the NGS mode of Gen\ II
in the I band under turbulence  conditions of 
$\omega_0=1.54,~t_0=2ms~at~0.88\mu m$.
The secondary component has been artificially rescaled
to the peak level of the primary. Zero position corresponds to the 
photocenter in the R band.  
On the left, contour levels are  1\%,5\%,10\%,30\%,50\% and 100\%  
and the magnitude difference of the binary is 2.5. 
Anisoplanaticism effect is seen on the  $2\farcs 42$ binary.  The elongation 
of the  secondary is clearly visible.  On the right, 
both PSFs look identical within the uncertainties. 
The magnitude difference is 1.5 and the separation is $1\farcs 75$.  
Contour levels are  10\%,30\%,50\% and 100\%.
 See text.}
\end{figure}

{\bf K band} 
From COP data, anisoplanaticism effects have not been clearly detected up to 
$13\farcs$ in K under various turbulence conditions.

{\bf 5.1.2 Temporal anisoplanaticism} 

Temporal anisoplanaticism will elongate the on-axis PSF in the 
 direction of the wind in the dominant
turbulent layer (Wilson and Jenkins 1994). We miss this information
to carry on any study on the data.

{\bf 5.2 Noise propagation in the wavefront sensor}

We have seen that it is possible to 
match the flux of the PSF calibrator and the astronomical target but
the wavefront sensor noise is still 
dependent of the source shape. For instance, 
for a binary, the centroide measurement will be noisier along its axis
and the on-axis PSF may be more elongated in this direction.
Let us quote the relevant parameters : the seeing angle $\omega_0$, 
the diffraction-limit of the sub-aperture $\omega_{sub}$.  For COP, 
we have typically  $\omega_0 = 1\farcs,~\omega_{sub} = 0\farcs24$.
For Gen~II, we have typically 
$\omega_0 = 1\farcs,~\omega_{sub} = 1\farcs5$. Such effects are expected to  
be negligible for sub-arcsecond binaries and quite difficult to detect on 
images.  In the case of wider binaries, the pixel width, the field of
view per subaperture and the flux ratio are important parameters. 
Gen~II uses only 2 by 2 pixels per subaperture with a pixel width of about 
$2\farcs4$. In the case of the $2\farcs42$ binary (see Fig. 7), no effects are 
visible on the on-axis PSF.

\begin{center}
{\large\bf 6. RESIDUALS FROM DECONVOLUTION AND PERFORMANCES}
\end{center}

Residuals and performances  presented in this section assume the use of 
classic  deconvolution  with the  PSF estimated from the observation of
 a point source (see 1.4.1). The typical COP observation of the closest binary 
($0\farcs 13$)  has been used again  to present the results of the subsection
 6.2  and 6.4.  The COP observation of a set of few binaries 
have been used  to present the results of the subsection 6.5.  
These results are  given as an example and do not be interpreted as 
the ultimate performances of this system since they depend  
on the seeing stability, the quality of the  calibration,  and the 
observational procedure during the observations; the two last points
can be improved. 

{\bf 6.1 A typical COP observation}

Table 2 describes the observations and Fig. 5 (top) shows the SR of the binary
and its calibrator.  
Short exposure  images were coadded and background corrected to get
the source image and the PSF image. DAOPHOT was used to computed the binary 
parameters from the  source image and the PSF image. A residual image is 
consequently available as the  substraction of the fitted binary convolved 
with  the PSF to the source  image.  

\begin{table*}
\caption{\label{tab:two}
Journal of a  COP observation}
\begin{flushleft}
\begin{tabular}{lllll}
Source & Band  & Observation time (UT)& Integration time (s)
& Exposure time (s)\\
\hline
Binary 		& J& 8:18 & 300 &1/2\\
       		& H& 8:31 & 300 &1\\
       		& K& 8:42 & 300 &60\\
Calibrator 	& K& 8:54 & 120  &1\\
		& H& 8:59 & 60  &1\\
  		& J& 9:03 &60   &20\\
\end{tabular}
\end{flushleft}
\end{table*}

{\bf 6.2 Seeing variations as a source of the deconvolution residuals}

We have seen in section 4 that the PSF is not stable.  
From the cube of continous PSF frames,  we have
 derived a statistic  of the PSF during the observation
which shows  that the signal to noise of the PSF
does not increase as the square root of time (elementary time is 
selected to a few seconds to get uncorrelated wavefront on the mirror from
frames to frames). We might not expect  a such signal to noise behavior if 
the variations of the PSF  were only induced e.g. by high Zernike 
uncorrected terms and the  wavefront sensor noise at constant seeing.  
The explanation is that the seeing  variations  
make the PSF as a non-stationary process.
Since it is not possible to describe the behavior of the signal to noise
for a non-stationary process, we will consider the  standard deviation  of the 
cube of PSF frames divided by the square root  of  the number of frames 
minus one  as the low  estimation  of the PSF noise. 

Let us have a look at the residuals from the 
deconvolution of the binary. The shape of the residuals is a noisy halo.
The plot of the histogram of the halo shows that it is in this case
systematically positively bias. The bias
in comparison with the width of the histogram is slight in HK  
but strong in J. It could be explained by an evolution of the 
seeing  (see Table 2) or by  a systematic lower correction on the
source due for example to a different flux on the WFS despite the precautions
taken during the observations.  Fig. 8 shows the profile (averaged 
on the pixels
located at the same distance to the peak maximum of the image) for
the absolute residuals and for the standard deviation of the PSF i.e. the
low estimation of the PSF noise.  Of course, both profiles
decrease as the radial distance. The PSF noise increases as the 
Strehl ratio drops.  Residuals noise is systematically above the PSF 
standard deviation.  The PSF mismatch for the calibrator and the source
is  responsible  for the level of  residuals.   
In J, the calibrator was pointed too much time 
 after the binary source (45  minutes) and deconvolved image reveals  
a strong residual halo.  In this case, seeing variation is likely to explain
a such effect. At K, the calibrator   was observed 12 minutes after 
the source and we  can see (Fig. 8) that  the residuals are lower but still
 up to five times larger than the PSF  standard deviation. 
Incidently, the first airy ring is a  sensitive area, it may happen 
to get artifact in the deconvolution   process up to 10\% of the 
peak level in the worst cases. 

To highlight the residuals due to the seeing variations rather than
some mismatch flux on the WFS, we have deconvolved
 the PSF calibrator  by the PSF calibrator taken a few  minutes apart. The
residuals  show again the same level than in Fig. 8. 
However, the residuals in J on Fig. 8 are certainly  a high estimation. 
At last, to estimate the contribution of the flat-fielding error, we used a 
PSF taken a few  minutes later but located on a different part of the 
detector,  residuals are not significantly different. All this confirms 
that the seeing variations are the source of strong residuals and the
  quality of the calibration  directly affects the residuals  level.

\begin{figure}
\psfig{height=8cm,angle=-90,width=16cm,file=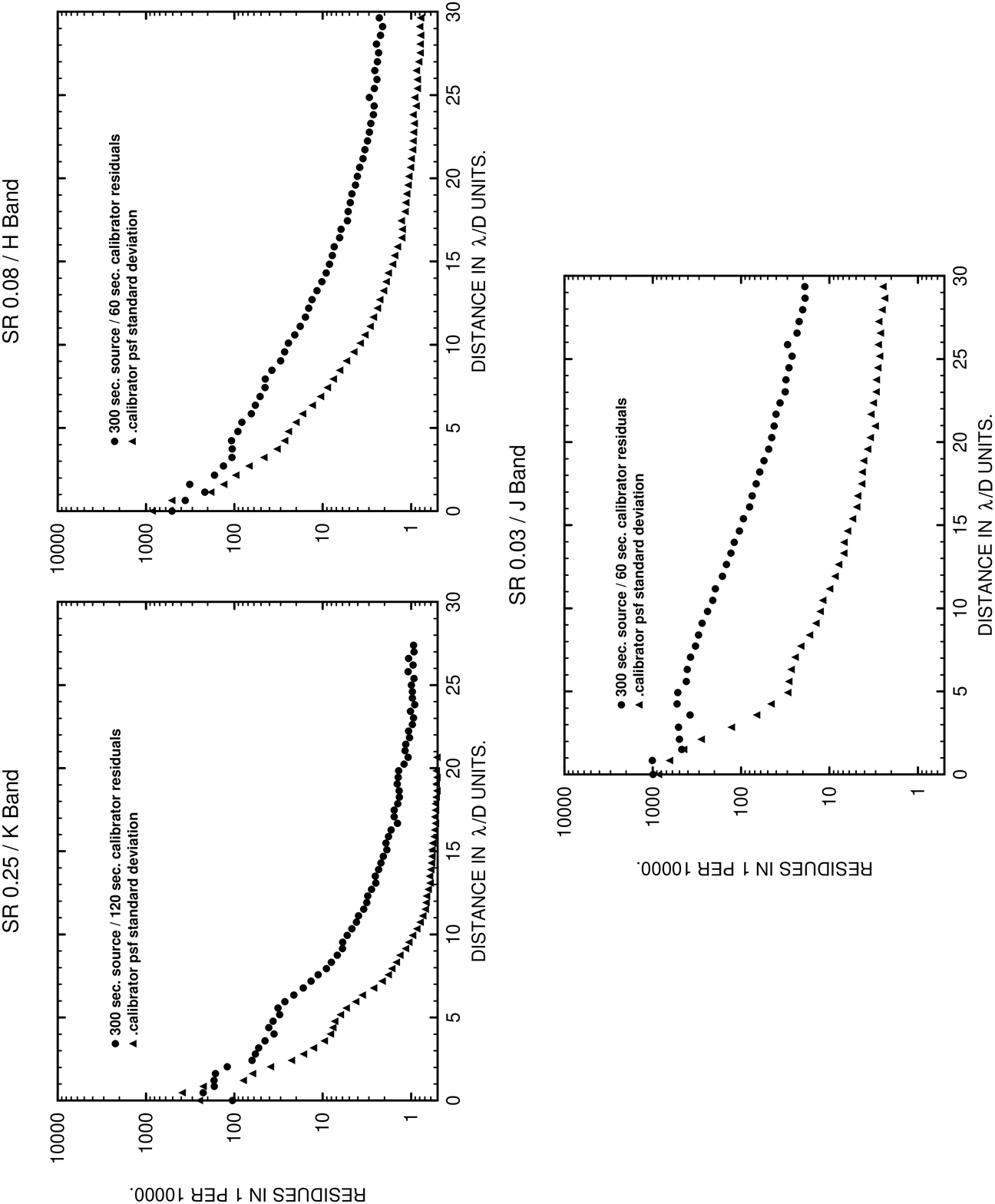}
\caption{The absolute residuals from deconvolution are here compared 
with the standard deviation of the PSF calibrator for different
Strehl ratios. From COP data. See text.}
\end{figure}

{\bf 6.3 Procedure to get a good quality  PSF calibration}

Observers cannot control the seeing variations but they can adjust the 
observation procedure so as to minimize its effets. Here are the 
few rules to follow.

\noindent i) Choose a PSF calibrator close to the astronomical target

ii) Match its flux to the astronomical target (use density filters e.g.).

iii) Observe the PSF calibrator longer enough  (one or two minutes)
to smooth the short timescale
seeing variations and the residual seeing halo as well.

iv) Observe the PSF calibrator as shortly as possible before or after 
the source  observations to avoid long-time scale seeing variations or the
seeing variation with the elevation.

v) If possible, repeat several times this alternate observation procedure.

These constraints are quite difficult to fullfill for faint 
astronomical targets which needs very long exposure times. However, in the 
infrared, because of the  background variations, exposure times are limited in
general to five minutes. 

\pagebreak
{\bf 6.4 Sensitivity curves}

{\bf Detection of a companion star}

We use the residuals to compute the typical level of detection for a companion.
This curve (Fig. 9) could be compared to two results from COP. First, 
the observation of the R~136 cluster has revealed stars in the field 
($13\farcs$) as 
faint as a magnitude difference of 9 relatively to the brightest star in the 
field (See Brandl et al.  1995). Secondly,  a faint companion in K with 
a flux ratio of $10^4$ 
at $4\farcs$ have been detected around the object HR4796 (See L\'ena 1994).  

\begin{figure}
\psfig{height=8cm,angle=-90,width=16cm,file=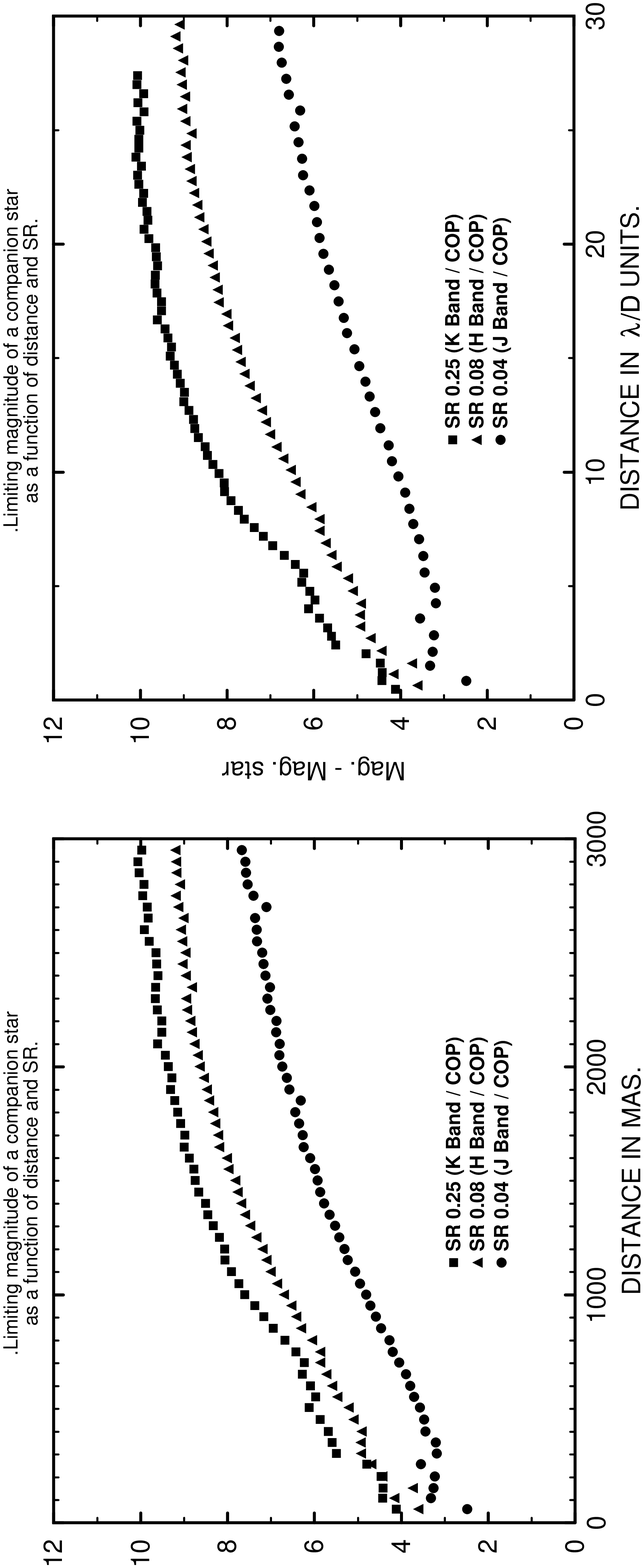}
\caption{COP sensisitivy curve for the detection of a companion in
magnitude difference to the main component as a function of the
separation for different Strehl ratios. Left
plot  in milli-arcseconds, the right one is rescaled by $\lambda/D$.
A pixel width of 50\ mas is assumed. See text.}
\end{figure}

{\bf Detection of an extended source around a point source}

We define an extended source as a source much larger than the
50\% energy radius of the PSF. Sensitivity curve are shown on Fig. 10
with a pixel of 50\ mas. Binning  data will gain in sensitivity as the 
binning factor.   The detection of the disk around $\beta$ Pictoris (A.-M.
Lagrange, private communication) at $2\farcs$ is consistent with this curve.

\begin{figure}
\psfig{height=8cm,angle=-90,width=16cm,file=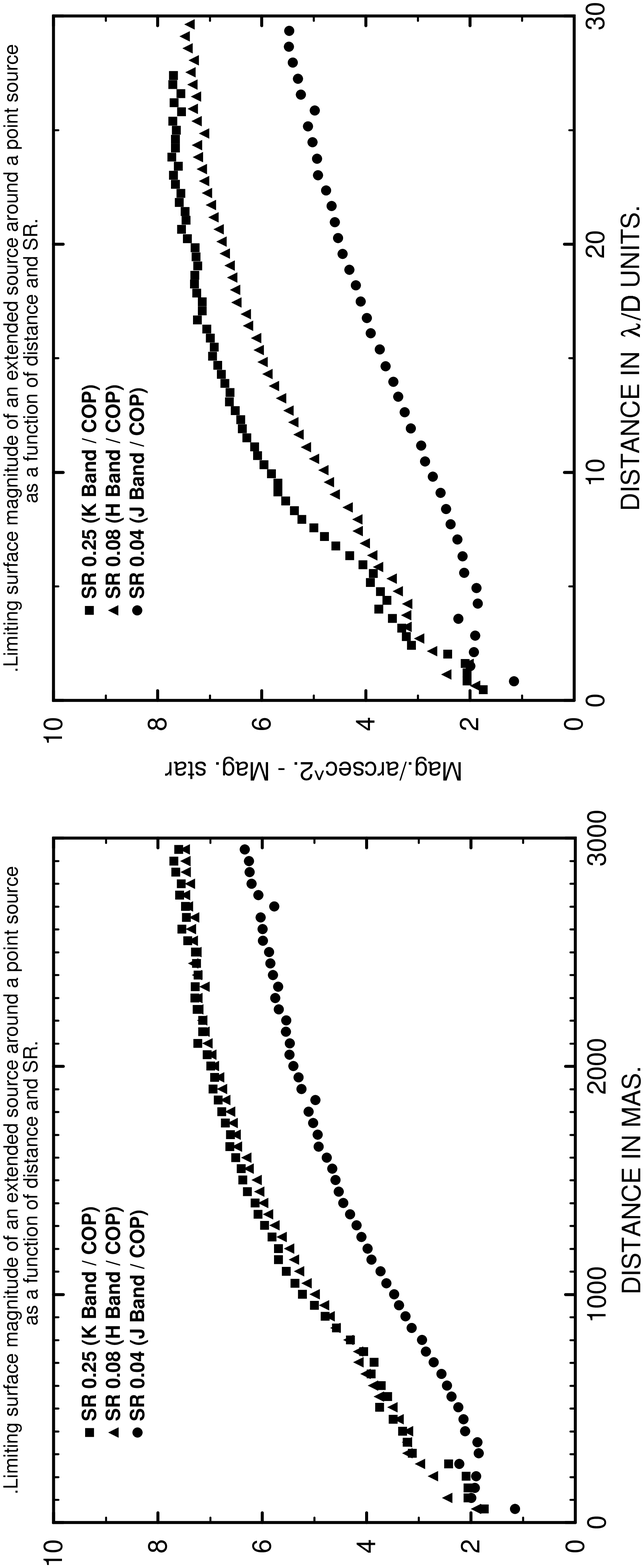}
\caption{COP 
sensisitivy curve for the detection of an extended structure around
a point source in magnitude per arc-second$^2$ relative 
to the central source as a function of the radial distance 
for different Strehl ratios. 
Left plot  in milli-arcseconds,  the right one is rescaled by $\lambda/D$.
A pixel width of 50\ mas is assumed. See text.}
\end{figure}

{\bf 6.4 Photometric and astrometry performances}

Residuals are an indication but it is not straightforward
to derive photometry and astrometry accuracy. COP data of a few
sub-arcsecond  binaries were divided in subsets  of one\ minute exposure 
time to derive statistical errors for one\ minute
exposure AO images.  DAOPHOT package have been used again 
to extract the  sub-arcsecond binary   parameters  from the source and 
calibrator images. Let first review the expected problems.

{\bf PSF variation due to seeing}
A mismatch between the PSF in the source and the calibrator yields
a noise in the photometry and the astrometry. The noise could be in both way 
and  increases with the PSF mismatch.

{\bf PSF variation due to anisoplanaticism}
SR fall off yields a under-estimated photometry for the secondary.  
This bias is proportional to the fall off. We have seen that for 
infrared AO data of sub-arcsecond binaries, this is not relevant.

Binaries are bright enough so that  the other sources
of noise in the images (photon, detector and background noises) 
are negligible in front of the deconvolution noise due to the
mitmatch of the PSF.  Again, this results  are partial and strongly 
depends on the quality of the calibration. A more complete study is under way.
Anyway, they show how the  accuracy increases  with the Strehl ratio e.g. 
The possible presence of a systematic  bias still remains to be investigated. 

{\bf Photometry }

Table 3 shows the flux error of the secondary component divided by the 
total flux of the binary system 
as a function of the separation with a  diffraction-limited
size of $0\farcs13$ and for a poor correction.  
Because of the seeing halo and a pretty low SR, 
photometry accuracy does not increase very rapidly with the angular
separation.

\begin{table*}
\caption{\label{tab:three}
Photometry accuracy as a function of separation (COP instrument)}
\begin{flushleft}
\begin{tabular}{llll}
Strehl ratio 8\% and $\lambda/D = 0\farcs13$\\
\hline
Separation(mas) & 200 & 500 & 700\\
Error (\%) & 3 & 1.5 & 1\\
\end{tabular}
\end{flushleft}
\end{table*}

Table 4 shows the flux error of the secondary component divided by the 
total flux for  a very close binary system (separation of $0\farcs13$)
as a function of the Strehl ratio  and the band. 
Despite the lower resolution in 
the K band, the higher SR improves the accuracy in the
photometry. 

\begin{table*}
\caption{\label{tab:four}
Photometry accuracy as a function of the Strehl ratio (COP instrument)}
\begin{flushleft}
\begin{tabular}{llll}
$\lambda/D$ & $0\farcs072$ & $0\farcs095$ & $\farcs13$ \\
\hline
Strehl ratio (\%)& 3 & 8 & 26\\
Error (\%) & 7 & 3 & 1.5\\
\end{tabular}
\end{flushleft}
\end{table*}

{\bf Astrometry}

Absolute astrometry relates to the abilility to superimpose  
AO images taken in different bands. For COP, the limitation is about
$20~mas$ and seems to be related to the instrumental rigidity (L\'ena, 
private communication). The relative astrometry accuracy (e.g. separation
of a binary) is higher and vary from 1 to a few mas.

\begin{center}
{\large\bf 7. PROSPECTIVE  MODES IN  AO}
\end{center}

{\bf 7.1 Selection and rebinning}  

Depending where the cloud of points is located on the curve
shown in Fig 1, we can estimated what the result of a selection will be. 
For instance, it is clear that a selection based on the FWHM is more 
worth in H than a selection
in K where the FWHM is well stabilized.   Selection 
is a efficient way to get sharper images. However for an resolved
object, the Strehl ratio is affected (see Fig. 5), as the others parameters
which can be used for a selection; there is no way to 
apply a similar selection for the source and its calibrator, consequently
 side effects from selection might strength miscalibration. 
The object-independent  estimation of the instanteneous Strehl ratio 
from   the WFS data   will be very helpful to apply some
selection or rebinning-like algorithms as in the speckle technique. 
 
{\bf 7.2 Speckle with Adaptive Optics}

We used here 250 milliseconds and 200 milliseconds frames taken with COP in JK 
respectively  under a seeing  of $1\farcs8$. Figure 9 compares  the 
theoritical speckle  transfer function with no AO correction derived from 
the $D/r_0$ (given by the estimated  seeing angle) to  the modulation 
transfer  function of  the  A0 frames 
($< \vert \tilde I \vert^2>$), of the long-exposure AO image 
($\vert < \tilde I >   \vert^2$ and of the 
shift-and-add AO image ($\vert I_{SAA} \vert^2$).  
For  Strehl ratios of 8\% in K and 1.3\% in J, 
short exposure images allows to increase
the signal in a large range of spatial frequencies. 
We miss the coherence time value to
fully compare  the speckle interferometry technique with or without AO but
certainly the coherence time is not greater than the exposure times used
here.  Consequently, speckle interferometry with AO leads substantial
gain.  Let us note that 
the signal to noise of the AO MTF as in speckle interferometry
 is one per frame in the high spatial  frequencies  range. Above a SR 
of 25\%, short exposure 
images do not yield significant gain. These results are consistent with the 
simulation from  Conan (1995).  The conclusion is that the use of short 
exposures could be beneficial  (access to high spatial frequencies) 
in case of the observations of a 
bright object under a  poor AO correction. However, 
the calibration  of the speckle transfer function  with AO correction 
remains a problem not yet investigated. Algorithms like triple  correlation, 
Knox-and-Thompson   could be tried  as well.

\begin{figure}
\psfig{height=8cm,angle=-90,width=16cm,file=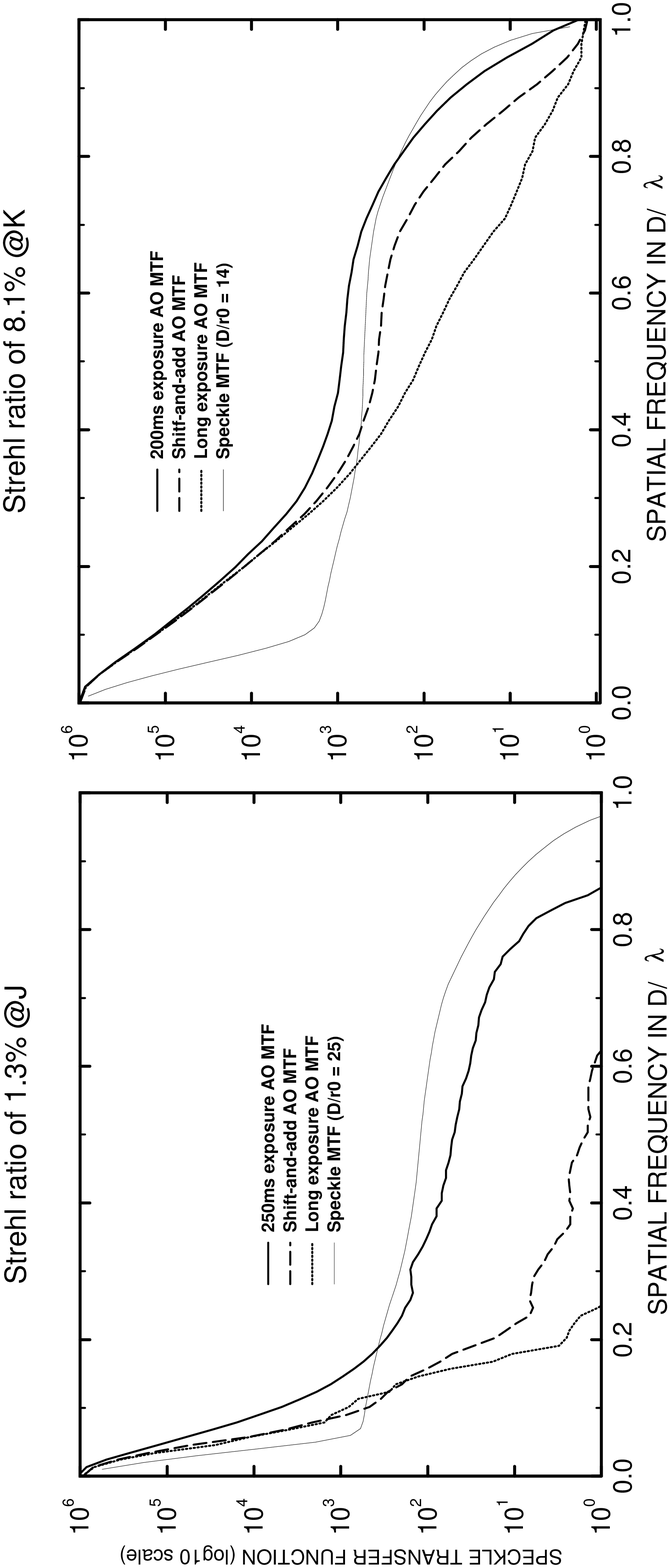}
\caption{Modulation transfer function of short exposures AO images (0.2s
and 0.25s in KJ respectively) under  poor correction  
compared to those of the long-exposure image and  the Shift-and-Add image,
and to the speckle transfer function with no AO correction. From 
COP data. See text.}
\end{figure}

\begin{center}
{\bf ANNEXE }
\end{center}

{\bf A1 How to compute the Strehl ratio of AO images?} 

We assume a  theoritical diffraction image obtained from a telescope
with a  linear central obstruction U and sampled by a  Shah function. 
Let define O as the ratio of the  Nyquist sampling to the 
Shah sampling.
 At constant flux set to 1.0, the maximum discrete value $M_0$ 
in this image is given by the formula: 

\begin{equation}
\label{eq:2}
M_0 = \frac{\pi}{16}\ O^{-2}\ (1-U^2) 
\end{equation}

Let consider an AO image obtained by a $N \times N$ array of detectors. 
We first normalize the flux to $1.0$ and then defilter from the  pixel function
(for a square pixel with a filling factor g, we divide the Discrete 
Fourier Transform of the image by the function $sinc (g^{1/2}\nu_x/N)
 sinc(g^{1/2}\nu_y/N)$). The maximum position of the image is located from  
a spline function interpolation.  The maximum M is deduced from the 
shifted image  via a Fourier transform to this position (equivalently, M is
equal to the integral of the Fourier transform recentered to this position).
 Let us note that the defiltering operation does not change the value of M
very much in case of low Strehl ratio images or oversampling 
data ($O \geq 1.5$). The Strehl ratio is then given by: 

\begin{equation}
\label{eq:3}
SR = M\ \frac{16}{\pi}\ O^{2}\ \frac{1}{1-U^2} 
\end{equation}

\pagebreak
\begin{center}
{\large\bf ACKNOWLEDGMENTS}
\end{center}

C. Perrier and J. Bouvier are gratefully acknowledged for providing 
COP data. I thank B. Ellerbroek and J. Christou for providing Gen~II data.
E. Tessier was supported by an European Union fellowship grant.

\begin{center}
{\large\bf REFERENCES}
\end{center}

1.  Beuzit J.-L., Hubin N., Gendron E., Demailly L., Gigan P., 
Lacombe F., Chazallet F., Rabaud D., Rousset G., 1994, Proc. SPIE 2201, p955

2. Brandl B., Sams B. and Eckart A.,  The messenger, March 1995.

3. Christou J.-C., Hege E.-K., Jefferies S.-M., Keller C.-U., 1994,
Proc. SPIE 2200, p433

4. Conan J.-M., 1995, Thesis, ONERA 

5. Fugate R., Ellerbroek B., Higgind c., Jelonek, Lange W., Slavin A.,
Wild W., Winker J., 
Spinhirne J., Boeke B., Ruane R., 
Moroney J., Oliker D., Swindle D., Cleis R., 1994, JOSA, Vol. 11, 1, p310 

6. L\'ena P., 1994, Proc. SPIE 2201, p1099

7. Tessier E., Bouvier J., Beuzit J.-L. and Brandner W., The messenger, 
Dec 1994. 

8. Thi\'ebaut E. \& Conan J.-M., 1994, JOSA in press

9. Wilson R. and Jenkins C., 1994,  MNRAS in preparation 

\end{document}